# Phase Coherent Precessional Magnetization Reversal in Microscopic Spin Valve Elements

*H.W. Schumacher[1], C. Chappert[1], P. Crozat[1], R.C. Sousa[2], P.P. Freitas[2], J. Miltat[3], J. Fassbender[4], and B. Hillebrands[4]*

[1] Institut d'Electronique Fondamentale, UMR 8622, CNRS, Université Paris Sud, Bât. 220, 91405 Orsay, France

[2] Instituto de Engenharia de Sistemas e Computadores, Rua Alves Redol, 9, 1º Dt., P-1000 Lisboa, Portugal

[3] Laboratoire de Physique des Solides, CNRS, Université Paris Sud, Bât. 510, 91405 Orsay, France

[4] Fachbereich Physik, Universität Kaiserslautern, Erwin-Schrödinger-Straße 56, 67663 Kaiserslautern, Germany

**Abstract**

We study the precessional switching of the magnetization in microscopic spin valve cells induced by ultra short in-plane hard axis magnetic field pulses. Stable and highly efficient switching is monitored following pulses as short as 140 ps with energies down to 15 pJ. Multiple application of identical pulses reversibly toggles the cell's magnetization between the two easy directions. Variations of pulse duration and amplitude reveal alternating regimes of switching and non-switching corresponding to transitions from in-phase to out-of-phase excitations of the magnetic precession by the field pulse. In the low field limit damping becomes predominant and a relaxational reversal is found allowing switching by hard axis fields below the in-plane anisotropy field threshold.

**Corresponding Author:**

Hans Werner Schumacher, IEF, Université Paris Sud, Bât. 220, F-91405 Orsay, France, e-mail: schumach@ief.u-psud.fr, phone: +33169154013, fax: +33169154000

**Article**

It is well known since the 1950s that the ultra fast dynamics of the magnetization in ferromagnets are governed by the damped precession of the magnetization about a local effective field (1). The use of the pronounced precession initiated by fast rising magnetic field pulses (2,3,4,5) is currently discussed as a new route towards ultra fast magnetization reversal in future magnetic storage applications. Such precessional reversal promises high energy efficiency together with ultra fast reversal times reaching the fundamental limit of a half precession period (6,7,8,9,10,11). In the first experimental demonstration of precessional switching intense magnetic field pulses of only a few ps duration were used to reverse large domains in Co and Co/Pt thin films (6,7). Numerical simulations pointed out that precessional switching should also be applicable for small magnetic cells (8,9) as used in magnetic random access memories (M-RAM) (12) and with technically available pulse durations longer than 100 ps. The first observations of such processes, in the limit of high pulse amplitudes and short pulse duration were recently reported (13). Here, we provide for the first time the full experimental evidence of the oscillating nature (8) of the precessional switching of the magnetization in microscopic memory cells. Ultra fast, stable, and reversible switching back and forth of the magnetization is shown to be triggered by hard axis pulses as short as 140 ps and with pulse energies down to 15 pJ. Variation of pulse duration and amplitude gives clear evidence for the dominating role of the phase coherence between the magnetic precession and the magnetic field pulse during reversal. At low pulse fields a transition to a damping dominated reversal is observed allowing switching by hard axis fields below the static threshold of the in-plane anisotropy field.

The experiments were carried out on $2 \times 4$ µm$^2$ exchange biased spin valves (SV) consisting of Ta 65Å / NiFe 40 Å / MnIr 80 Å / CoFe 43 Å / Cu 24 Å / CoFe 20 Å / NiFe 30 Å / Ta 8 Å with the exchange bias field and the magnetic easy axis of the soft layer along the long dimension. Fig. 1(a) shows a typical static easy axis field ($H_{easy}$) magneto resistance (*MR*) loop of the free layer of one of our devices. The loop is square on the left and exhibits kinks to the right as well as a clear asymmetry characteristic of exchange-biased devices (14). Due to the electrical contacts the *MR* measurement only probes the magnetization within the 3µm wide center region of the cell. Flux closure domains underneath the contacts thus do not significantly contribute to the loop. The loop is shifted to an offset field $H_{offset} \approx 14$ Oe due to coupling of the pinned and the free layer (15), and reveals a *MR* change of 5.6 % at $H_{easy}=H_{offset}$. The SV elements are integrated into a device comprising on-chip high-bandwidth pulse lines that generate magnetic field pulses upon current pulses injection (16). The transient pulses are monitored using a 50 GHz sampling oscilloscope. The pulse durations $T_{pulse}$ can be adjusted between 140 ps and 10 ns (at half maximum) with rise times down to 45 ps (from 10-90%) and maximum fields around 240 Oe. In addition, an external coil allows application of in-plane static fields.

In our experiment we follow theoretical predictions (8,9), suggesting that switching can be achieved by field pulses along the in-plane magnetic hard axis of a memory cell. The pulsed field is thus oriented at $\pm 90°$ angle from the initial or final magnetization direction and the switching is expected to be symmetric, i.e. the magnetization can switch forth and back by consecutive application of the same field pulse. Such a reversible precessional switching process is shown in Fig. 1(b).

First, the sample is saturated into the low resistance (parallel) state by applying $H_{easy}$ = -100 Oe before increasing $H_{easy}$ to $H_{offset}$ (15 Oe in Fig.1) to compensate the loop shift. The pulse response of the cell is then tested by consecutively applying identical hard axis pulses. In Fig 1(b) the *MR* value measured after each pulse is plotted versus pulse index for three different pulse parameters. For $T_{pulse}$ = 140 ps, $H_{pulse}$ = 155 Oe (solid circles) each pulse toggles the magnetization direction. Each *MR* change $\Delta MR$ of ±5.6 % corresponds to the full remanent reversal measured at $H_{offset}$ (cp. arrows in (a)). So far the zero error reliability of this reversible switching process has been tested for up to 750 consecutive pulses. For slightly longer pulses ($T_{pulse}$ = 190 ps, 195 Oe: gray squares) reversible switching no longer occurs. Here, the *MR* changes stochastically between 0.01 and 1.5 %. Finally, for 270 ps, 215 Oe pulses (open circles), the sample is always found in the parallel low resistance state, meaning that the initial and final states are essentially identical (no effective switching) in spite of the fact that the field strength was even slightly increased (18).

The damped precession of the magnetization vector *M* about the local effective field $H_{eff}$ is described by the Landau-Lifshitz-Gilbert equation (1)

$$\frac{dM}{dt} = -\gamma (M \times H_{eff}) + \frac{\alpha}{M_S}\left(M \times \frac{dM}{dt}\right)$$

with $\gamma$ being the gyromagnetic ratio, $\alpha$ the damping parameter and $M_S$ the saturation magnetization. From the first term on the right hand side we see that the torque on *M* is maximum when *M* and $H_{eff}$ are oriented perpendicular to each other. As sketched in Fig 2(a) such is the case right after the onset of the hard axis field $H_{pulse}$ and the torque $-\gamma(M \times H_{pulse})$ induces a rotation of *M* out of the film plane. This out-of-plane component, in turn, generates a strong demagnetizing field $H_D$ (17) also oriented perpendicular to the plane but pointing in the direction opposite to the $z$ magnetization component, as sketched in Fig 2(b). Under the action of $H_D$, $M$ remains mainly in-plane, and $H_D$ and $M$ are still oriented almost perpendicularly to each other allowing for fast quasi in-plane rotation of $M$ under the action of the new torque $-\gamma(M \times H_D)$. If the pulse terminates at about half the precession time, then a full switching is expected to be stabilized. Also the reversibility of the switching by multiple unipolar pulses is well accounted for by the model. Indeed, a reversal of $M$ also changes the sign of the dipolar field $H_D$ leaving the reversal mechanism unchanged.

These considerations can be substantiated by numerical simulations performed in the macro-spin (or single spin) approximation (8). In these simulations, the free layer is modeled by the value of $4\pi M_S = 10800$ Oe of permalloy and the demagnetization factors $N_X = 0$ (easy axis), $N_Y/4\pi = 0.0067$ (in-plane hard axis), $N_Z/4\pi = 0.9933$ (out of plane) chosen to mach the measured in plane anisotropy field of $H_A \approx 70$ Oe. The effective damping parameter (9) is chosen equal to $\alpha=0.03$ based on previous work (16). Lastly, the pulse shape mimics the transient current pulses as measured by the oscilloscope.

Fig. 2(c) displays the calculated normalized components of $M$ plotted as a function of time for the 140 ps duration pulse. The reversal is reproduced and the easy axis component $m_X$ changes sign. The non-zero value of the out-of-plane component $m_Z$ and thus the influence of the demagnetizing field $H_D$ are also clearly seen. Precession motion is exemplified by the $\pi/2$ phase shift between $m_X$ and $m_Z$. Furthermore, the fact that $M$ passes through an almost perfect alignment with the hard axis ($m_X$ and $m_Z \approx 0$, $m_Y \approx 1$) confirms the quasi in-plane rotation of *M*. Upon pulse termination *M* finally relaxes towards the nearest attractor, namely $m_X = 1$. The simulated non-reversal event induced by the 270 ps pulse is displayed in Fig. 2(d). As a consequence of the longer pulse duration $m_X$ now oscillates from the -*x* direction to the reversed orientation *and* back before the pulse terminates. After field switch-off, the magnetization again relaxes towards the closest attractor which now is $m_X = -1$. Therefore, in spite of the strong precessional motion of *M*, no effective cell reversal takes place. Simulating the observed partial switching ($T_{pulse} = 190$ ps) goes beyond the limits of a macro spin model, however, the weak reproducibility can be explained owing to the same picture. If the pulse decays when *M* is oriented near the hard axis i.e. near the energetic saddle point, a small trajectory variation may change the final attractor and thus induce a nearly stochastic reversal.

For a weak damping, a further increase of $T_{pulse}$ will lead to multiple oscillations of *M* about $H_{pulse}$. In the limit of strong pulses ($H_{pulse} \gg H_A$) and weak damping, switching is expected whenever the pulse terminates out of phase with the precession, i.e., when $T_{pulse} \approx (n+1/2) \cdot T_{prec}$, *n* being an integer defining the order of the switching process and $T_{prec}$ the time for one precession period. On the contrary, pulses parameters with $T_{pulse} \approx n \cdot T_{prec}$ will result in an effective non-reversal of the cell. Such phase coherent, higher order reversal is indeed observed experimentally as shown in Fig. 3 for a pulsed field strength $H_p \approx 230$ Oe (18). We plot the degree of reversible switching for a second, similar device in Fig. 3(a) as a function of $T_{pulse}$, and in Fig. 3(b) as a gray-scale map with $T_{pulse}$ as the *x*-coordinate and the reduced easy axis field $H_{red} = H_{easy} - H_{offset}$ as the *y*- coordinate. As a measure of reversible switching reliability we use $<|\Delta MR|>$, the absolute resistance change per applied pulse normalized to full reversal, and averaged over a series of pulses. $<|\Delta MR|> \approx 1$ indicates stable reversible switching.

Fig. 3(a) shows the measured values of $<|\Delta MR|>$ for $H_{red} = 0$. Four regions of stable switching, well separated by stable non-switching regions, are observed. Note again the zero order switching ($n = 0$) for the shortest accessible pulses of $T_{pulse} = 140$ ps. In the adjacent region up to $T_{pulse} \approx 280$ ps no switching takes place as $T_{pulse} \approx T_{prec}$. Additionally, higher order switching processes ($n = 1,2,3$) are observed near 350 ps, 590 ps and 800 ps. For $n = 1$ full, stable switching is obtained. We note, however, a slight decrease of $<|\Delta MR|>$ for higher reversal orders ($n = 2,3$) due to less reliable and partial switching. Note, that stable switching is observed over a relatively broad range of $T_{pulse}$ of about 100 ps indicating a large tolerable phase mismatch of the coherent switching still allowing stable attraction to the reversed easy direction (19). Also up to $H_{red} \approx \pm 4$ Oe (Fig 3(b)), corresponding to about 28% of the static loop coercivity ($H_C = 14$ Oe), the SV still reversiby switches ($n = 0,1$). However, easy axis fields $|H_{red}| \geq 0.5\ H_C$ finally result in irreversible switching, $M$ ending always aligned with $H_{red}$ whichever its initial orientation.

These results are of vital importance for M-RAM applications. The relatively broad pulse parameter range for coherent switching should e.g. allow down scaling of the cells. Switching of each cell of an array by a given pulse seems possible despite an inevitable spread in cell parameters (e.g. due to shape variations). Moreover, bit addressing can be realized in the standard cross line architecture (12) using the irreversible switching at higher easy axis fields.

A characterization of the switching properties over a wide range of $H_{\text{pulse}}$ and $T_{\text{pulse}}$ is displayed in Fig. 4 for $H_{\text{red}} = 0$. $<|\Delta MR|>$ is found in the upper panel and the results of the macro spin simulation in the lower one. The simulated field dependence of the coherent regimes is in good agreement with the measurements, and the minimum switching field of about 50 Oe is well reproduced. A decrease of $H_{\text{pulse}}$ first shifts the switching regions towards larger values of $T_{\text{pulse}}$. This is expected from simple ferromagnetic resonance arguments (17), since the precession period generally increases with decreasing field. Interestingly enough, just above the switching field limit of about 50 Oe, simulations predicts $n = 0$ switching *independent* of $T_{\text{pulse}}$. Thus, the coherence criterion appears to be no longer valid. However, now the condition $H_{\text{pulse}} \gg H_A$ is no longer fulfilled. *M* will overcome the hard axis only once during pulse application and precess around an equilibrium direction defined by the ratio of $H_{\text{pulse}}$ and $H_A$ as seen in Fig. 2(e) for a 5 ns, 55 Oe pulse. Owing to the damping, oscillation back across the in-plane hard magnetization axis is inhibited and $m_X$ changes sign only once. Thus, independently of $T_{\text{pulse}}$, switching will be completed through relaxation towards $m_X = +1$ once the pulsed field decays. Note again, that $H_{\text{pulse}} < H_A$ i.e. contrary to previous statements (13) the switching fields can be *lower* than the static hard axis switching threshold. Experiments also provide evidence for such damping dominated relaxational switching with $H_{\text{pulse}} \approx$ 50 Oe $< H_A$ for $T_{\text{pulse}} = 300...800$ ps. However, $<|\Delta MR|>$ remains well below 1 indicating only partial and instable switching. A reason for this is that for the given low $\alpha$ this switching mode is limited to a narrow field range. Weak parameter inhomogeneity, due e.g. to magnetic microstructure, might already be sufficient to inhibit the full switching of our cells. Smaller cells with a more single-domain-like behavior and a larger $\alpha$ might help stabilizing this switching mode. Note, that conversely, the coherent reversal regimes only weakly depend on $\alpha$, at least for $n=0$. Lastly, with further reduction of $H_{pulse}$ the precession amplitude becomes too weak to reach the in-plane hard axis and reversal is inhibited. Fig. 4. also allows to find parameters for optimal switching efficiency. Switching is obtained by pulses of $T_{pulse} = 140$ ps and only 15 pJ energy (arrow) proving high efficiency when compared to conventional easy axis switching schemes (20,21).

Concluding, we have studied the precessional switching of magnetic memory cells as conceptually used in M-RAMs. Ultra fast, reversible switching by hard axis pulses as short as 140 ps and with pulse energies down to 15 pJ was achieved. The switching behavior was fully linked to the characteristic times of magnetization precession. The existence of a phase coherent reversal regime at high fields, and a damping dominated, relaxational regime at low fields was established. In the relaxational regime, switching by pulse fields below the in-plane anisotropy field i.e. below the static hard axis field threshold was demonstrated. (22)

**FIGURE CAPTIONS:**

**FIGURE 1:**

Precessional reversal of the magnetization of a 2 μm × 4 μm spin valve cell. (a) static easy axis hysteresis loop. Magneto resistance *MR vs.* easy axis field. The loop is shifted by $H_{offset}$ = 14 Oe. The relative magneto resistance variation $\Delta MR$ at $H_{offset}$ is 5.6 %. (b) precessional magnetization reversal by hard axis field pulses. *MR* measured after each pulse *vs.* pulse index. $H_{offset}$ is compensated. Pulse parameters are $T_{pulse}$ = 140 ps, $H_{pulse}$ = 155 Oe (black, full circles); 190 ps, 195 Oe (gray squares); and 270 ps, 215 Oe (gray, open circles). The 140 ps hard axis pulse reversibly toggles the free layer magnetization with every pulse. $\Delta MR$ = 5.6% per pulse corresponds to a full cell reversal.

**FIGURE 2:**

Precessional switching by hard axis pulses. Sketches: (a) The magnetization *M* is first tilted out of plane by $H_{pulse}$, and then (b) rotates about $H_D$ towards the new direction -*M*. (c-e) Macrospin simulations of the hard axis pulse response. Normalized components of *M*=($m_X$, $m_Y$, $m_Z$) as a function of time. The scale for $m_Z$ is increased by a factor of 5 for clarity. The gray background represents the pulse time evolution in arbitrary units. (c) reversal for a 140 ps duration pulse. (d) non-reversal for 270 ps duration pulse. *M* oscillates beyond the hard axis *and* back during pulse application. (e) relaxational reversal for a 5 ns, 55 Oe pulse. *M* overcomes the hard axis only once during pulse application. Oscillation back is inhibited due to damping.

**FIGURE 3:**

Switching behavior of a second 2 µm × 4 µm spin valve cell as a function of pulse duration $T_{pulse}$ and reduced easy axis field $H_{red} = H_{easy} - H_{offset}$. Nominal pulse field is $H_{pulse} = 230$ Oe (18). (a) $<|\Delta MR|>$ the normalized average of $|\Delta MR|$ per pulse as a function of $T_{pulse}$ at exact loop offset compensation $H_{easy} = H_{offset} = 12$ Oe. Four regions of reversible hard axis switching around $T_{pulse} = 140, 350, 590$ and $800$ ps are found. (b) gray scale map of $<|\Delta MR|>$ as a function of $T_{pulse}$ and reduced easy axis field $H_{red}$. Gray scale legend is given in (a). black: $<|\Delta MR|> \leq 0.1$, no reversible switching; white: $<|\Delta MR|> \geq 0.9$ stable, large amplitude reversible switching, gray: intermediate values, instable or low amplitude switching. (a) corresponds to a section of (b) along $H_{red} = 0$.

**FIGURE 4:**

Pulse field dependence of the precessional switching of the SV cell shown in Fig.3. (a) $<|\Delta MR|>$ as a function of $T_{pulse}$ and $H_{pulse}$. $H_{red} = 0$ Oe. gray: $<|\Delta MR|> \leq 0.1$, white: $<|\Delta MR|> \geq 0.8$. $H_{pulse}$ is attenuated in steps of 1 dB. Switch pulse energies are as low as 15 pJ (arrow). (b) calculated switching map for the same device derived from macro spin simulations. black: no switching, white: switching. The order $n$ of switching is indicated.

**References:**

(1) L. Landau, and E. Lifshitz, Phys. Z. Sowjetunion **8**, 153, (1935); T. L. Gilbert, Phys. Rev. **100**, 1243 (1955).

(2) W. K. Hiebert, A. Stankiewicz, and M. R. Freeman, Phys. Rev. Lett. **79**, 1134 (1997).

(3) Y. Acreman et al., Science **290**, 492 (2000).

(4) T. M. Crawford et al., Appl. Phys. Lett. **74**, 3386 (1999).

(5) S. E. Russek, S. Kaka, and M. J. Donahue, J. Appl. Phys. **87**, 7070 (2000).

(6) C. H. Back et al., Phys. Rev. Lett., **81**, 3251 (1998)

(7) C. H. Back, et al., Science, **285**, 864 (1999).

(8) M. Bauer et al., Phys. Rev. B **61**, 3410 (2000).

(9) J. Miltat, G. Alburquerque, and A. Thiaville in *Spin Dynamics in Confined Magnetic Structures*, B. Hillebrands and K. Ounadjela (eds.) (Springer, Berlin, 2001)

(10) J. Miltat and A. Thiaville, Science **290**, 466 (2000).

(11) Y. Acreman et al., Appl. Phys. Lett. **79**, 2228 (2001)

(12) see e.g. S. Tehrani et al., IEEE Trans. Mag. **36**, 2752 (2000).

(13) S. Kaka and S. E. Russek, Appl. Phys. Lett. **80**, 2958 (2002).

(14) T. Ambrose and C. L. Chien, J. Appl. Phys. **83**, 7222 (1998)

(15) D. Wang et al., IEEE Trans. Magn. **36**, 2802 (2000).

(16) H. W. Schumacher et al., Appl. Phys. Lett. **80**, 3781 (2002).

(17) C. Kittel, *Introduction to Solid State Physics*, 5th ed., Wiley, New York, 1976.

(18) For a nominally constant pulse amplitude $H_{\text{pulse}}$ the measured $H_{\text{pulse}}$ is only constant for $T_{\text{pulse}} > 300$ ps. Below, it decreases almost linearly with $T_{\text{pulse}}$ down to 65% of the nominal value due to limitations of the pulse generator.

(19) A large phase mismatch will result in a pronounced precession ("ringing") of *M* upon pulse termination. However, in the center of the switching regimes only weak ringing due to a small phase mismatch and thus ultra short effective reversal times can be expected.

(20) R. H. Koch et al., Phys. Rev. Lett. **81**, 4512 (1998).

(21) B. C. Choi et al., Phys. Rev. Lett. **86**, 728 (2001).

(22) HWS acknowledges financial support by the European Union (EU) Marie Curie fellowship HPMFCT-2000-00540. The work was supported in part by the EU Training and Mobility of Researchers Program under Contract ERBFMRX-CT97-0147, and by a NEDO contract "Nanopatterned Magnets".

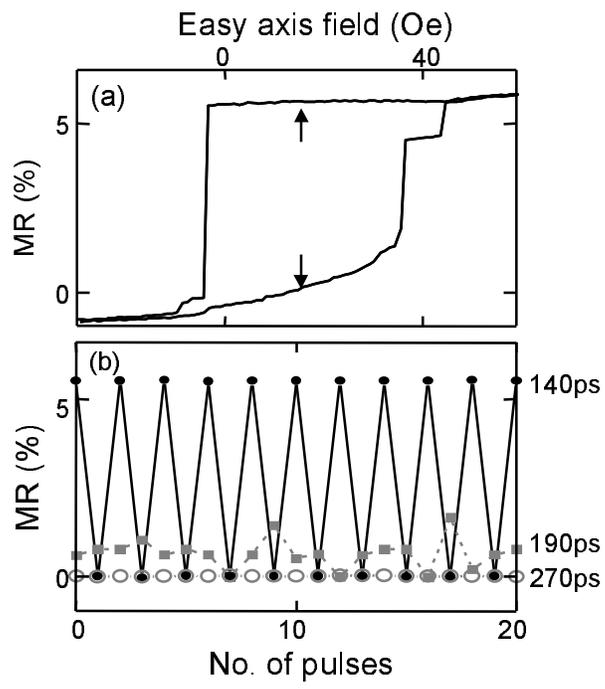

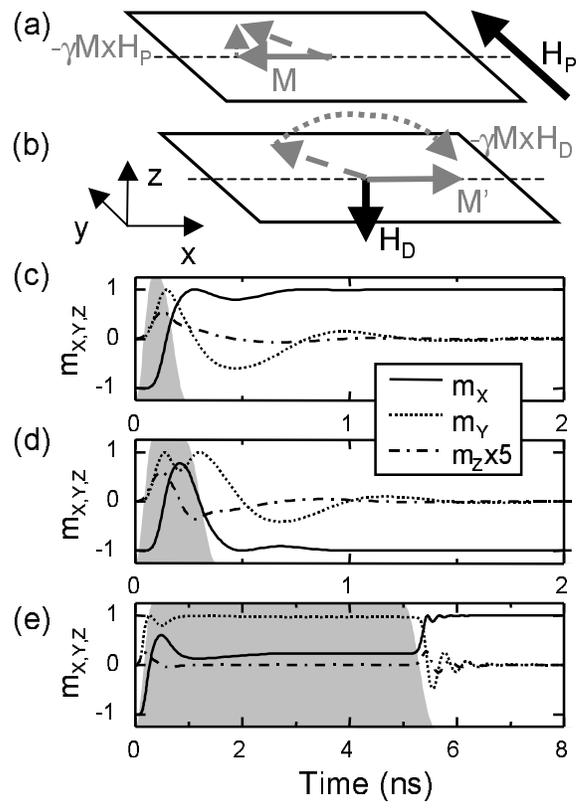

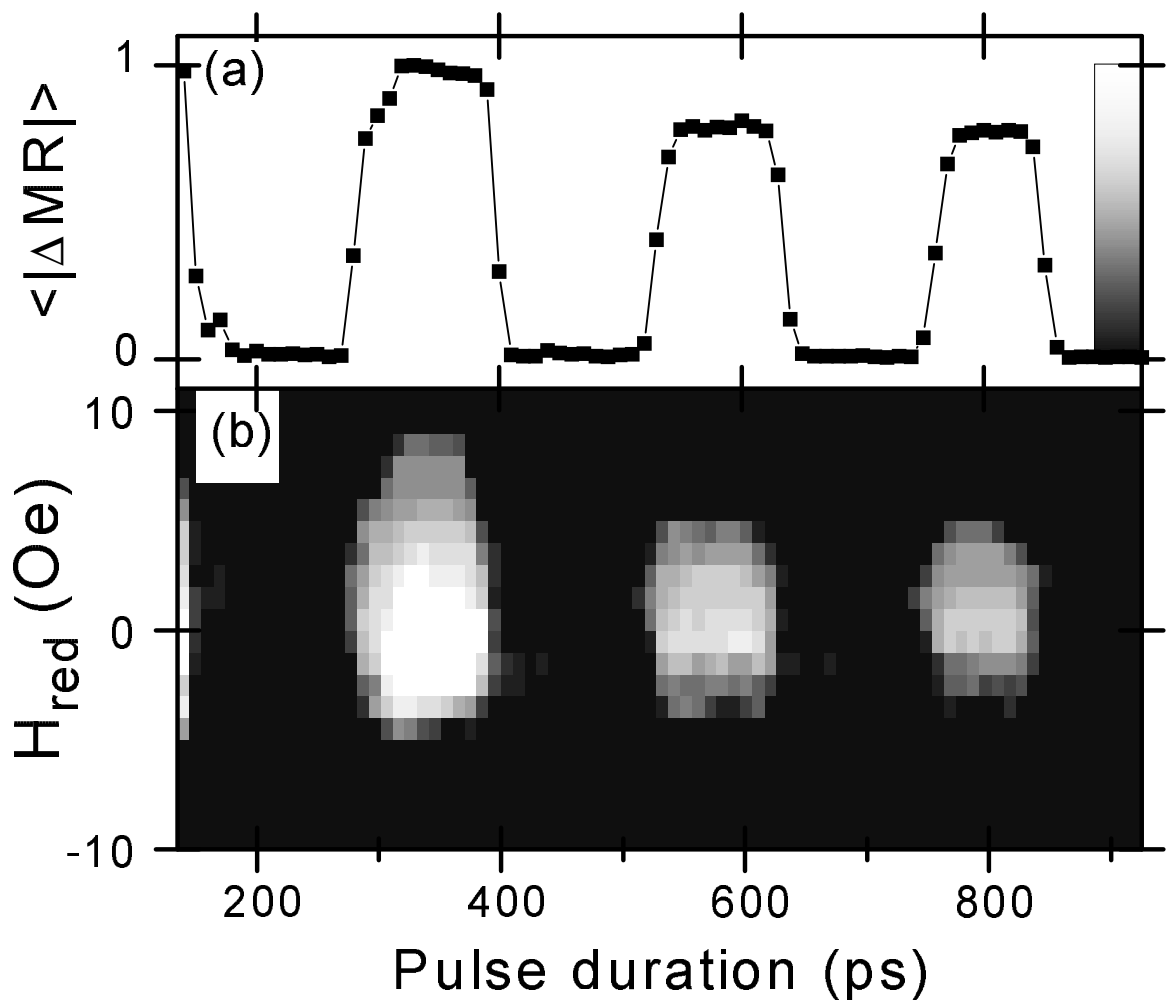

H.W. Schumacher et al.

Figure 3 of 4

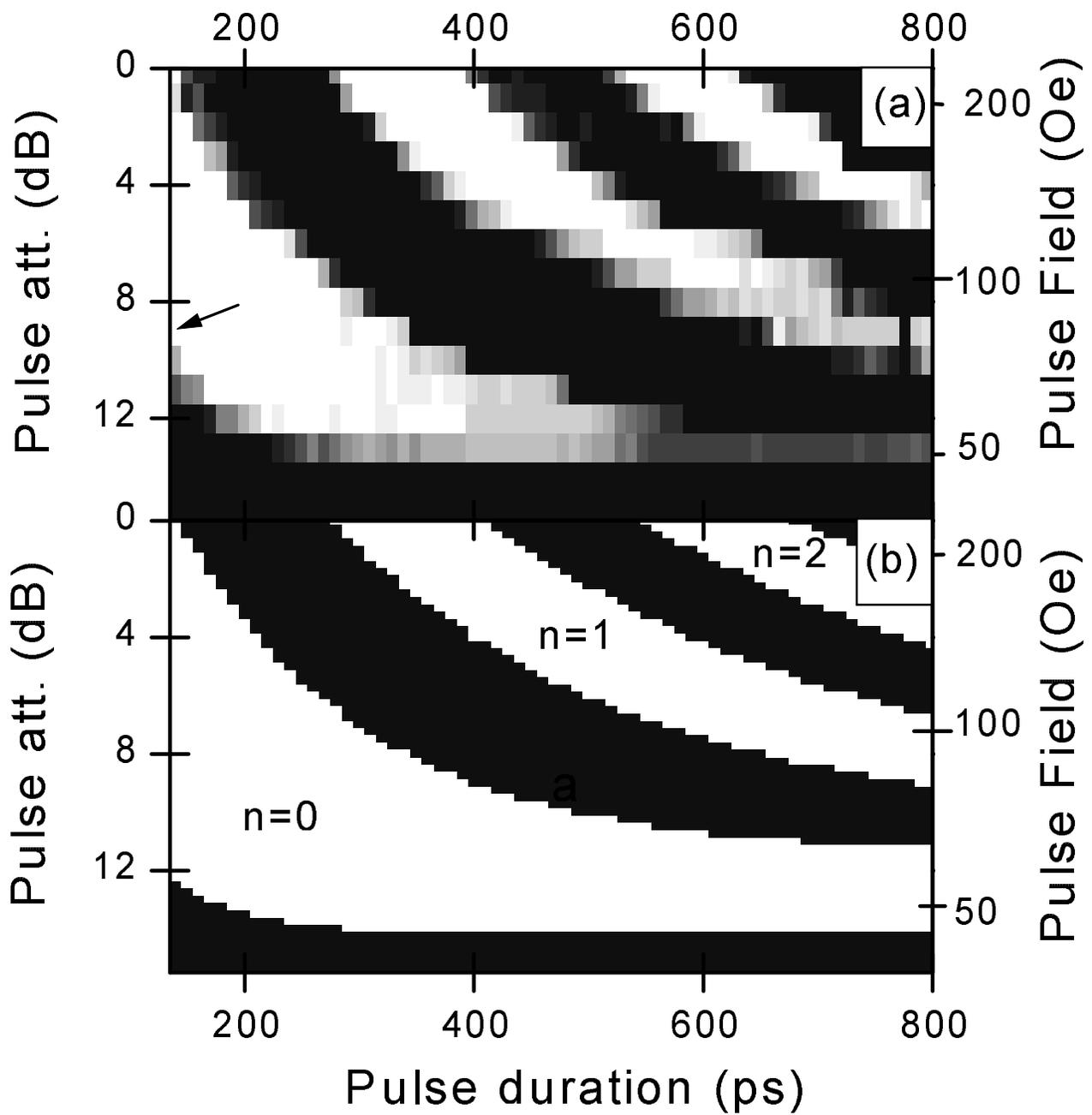

H.W. Schumacher et al.

Figure 4 of 4